\documentclass[aps,prc,reprint,groupedaddress]{revtex4-1}
\usepackage{graphicx} 
\usepackage{dcolumn} 
\usepackage{bm}
\usepackage{hyperref}

\begin{document}

\title{Nuclear fragmentation induced by low-energy antiprotons within a microscopic transport approach}
\author{Zhao-Qing Feng}
\email{Corresponding author: fengzhq@impcas.ac.cn}

\affiliation{Institute of Modern Physics, Chinese Academy of Sciences, Lanzhou 730000, People's Republic of China }

\date{\today}
\begin{abstract}
Within the framework of the Lanzhou quantum molecular dynamics (LQMD) transport model, the nuclear fragmentation induced by low-energy antiprotons has been investigated thoroughly. A coalescence approach is developed for constructing the primary fragments in phase space. The secondary decay process of the fragments is described by the well-known statistical code. It is found that the localized energy released in antibaryon-baryon annihilation is deposited in a nucleus mainly via pion-nucleon collisions, which leads to the emissions of pre-equilibrium particles, fission, evaporation of nucleons and light fragments etc. The strangeness exchange reactions dominate the hyperon production. The averaged mass loss increases with the mass number of target nucleus. A bump structure in the domain of intermediate mass for heavy targets appears owing to the contribution of fission fragments.

\begin{description}
\item[PACS number(s)]
 25.43.+t, 24.10.Lx, 25.70.Pq
\end{description}
\end{abstract}

\maketitle

\section{Introduction}

The properties of highly excited nuclei (hot nuclei) attract much attention in past decades via heavy-ion and hadron induced nuclear reactions, which are related to many interest issues, i.e., liquid-gas phase transition, fast fission, new isotope production, nuclear equation of state etc \cite{Ch04,Co93,Poch95,Wu98,Ma99}. The decay mechanism of the hot nuclei varies with the excitation energy. For example, particle evaporation and fission dominate the de-excitation process at the excitation energies of 1-2 MeV/nucleon, in which the nuclear structure effects are of importance, i.e., separation energy, shell effect, odd-even effect etc. At high excitation energies, the explosive decay is pronounced via the multifragment disintegration. The mechanism has been extensively investigated in heavy-ion collisions. The fragments are produced tend to the $\beta$-stability line. The symmetry energy plays an important role on rare isotope production. A large angular momentum transit and compression of nuclear system are undertaken in heavy-ion collisions, which complicate the formation of hot nuclei. To avoid the shortcomings, hadron-nucleus collisions could be chosen for heating nuclei. A more localized energy is deposited in a nucleus in antiproton induced reactions. For example, the average excitation energy after the stopped antiproton annihilation in Cu target is similar to the average excitation energy irradiated by 2-GeV proton \cite{Ja93}.

The decay mechanism of hot nuclei heated by stopped and energetic antiprotons has been investigated with the facility of the low-energy antiproton ring (LEAR) at CERN \cite{Eg90,Ho94,Ki96,Ja99,Lo01}. Some interest findings were reported, e.g., the delayed fission by antiproton annihilation in heavy nuclei \cite{Bo86}. The reaction dynamics induced by antiprotons is complicated and associated with the annihilation between antibaryon and baryon, charge-exchange reactions, elastic and inelastic collisions. To understand the nuclear dynamics induced by antiprotons, several approaches have been proposed, such as the intranuclear cascade (INC) model \cite{Cu89}, kinetic approach \cite{Ko87}, Giessen Boltzmann-Uehling-Uhlenbeck (GiBUU) transport model \cite{La09,Ga12}, Statistical Multifragmentation Model (SMM) \cite{Bo95} and the Lanzhou quantum molecular dynamics (LQMD) approach \cite{Fe14}. Antiproton-nucleus collisions provide the possibilities for studying the properties of hot nuclei, meson-nucleon interactions, formation of hypernucleus etc. It has advantage to heat the nucleus with the excitation energy of several hundreds of MeV and with less compression of nuclear system because of the annihilation reactions. In this work, the fragmentation mechanism in low-energy antiproton induced nuclear reactions is investigated, in which  the nuclear structure effects and reaction channels are discussed.

\section{Brief description of the model}

The LQMD transport model has been successfully used for the isospin physics in heavy-ion collisions, particle production and in-medium effects in heavy-ion and hadron ($p, \overline{p}, \pi, K$) induced reactions, hypernucleus production etc. In the model, the production of resonances, hyperons and mesons is coupled to hadron-hadron collisions, annihilation reactions of antibaryon-baryon collisions, decays of resonances, in-medium corrections on threshold energies, and transportation in mean-field potentials \cite{Fe11,Fe13}. The temporal evolutions of baryons (nucleons, resonances and hyperons), anti-baryons and mesons in the nuclear collisions are governed by Hamilton's equations of motion. The Hamiltonian of nucleons and nonstrangeness resonances is constructed within the Skyrme effective interaction, in which the isospin and momentum dependent potential is implemented. The Hamiltonian of hyperons ($\Lambda$, $\Sigma$ and $\Xi$) and pseudoscalar mesons ($\pi$, $\eta$, $K$ and $\overline{K}$) is derived from the relativistic covariant theories based on the fitting available optical potentials \cite{Fe13,Fe15}.

The mean-field potential of antinucleon is composed of the G-parity transformation of nucleon self-energies with a scaling approach. The antinucleon energy in nuclear medium is evaluated by the dispersion relation as
\begin{equation}
\omega_{\overline{N}}(\textbf{p}_{i},\rho_{i})=\sqrt{(m_{N}+\Sigma_{S}^{\overline{N}})^{2}+\textbf{p}_{i}^{2}} + \Sigma_{V}^{\overline{N}}
\end{equation}
with $\Sigma_{S}^{\overline{N}}=\Sigma_{S}^{N}$ and $\Sigma_{V}^{\overline{N}}=-\Sigma_{V}^{N}$.
The nuclear scalar $\Sigma_{S}^{N}$ and vector $\Sigma_{V}^{N}$ self-energies are computed from the well-known relativistic mean-field model with the NL3 parameter ($g_{\sigma N}^{2}$=80.8, $g_{\omega N}^{2}$=155 and $g_{\rho N}^{2}$=20). The optical potential of baryon or antibaryon is derived from the in-medium energy as $V_{opt}(\textbf{p},\rho)=\omega(\textbf{p},\rho)-\sqrt{\textbf{p}^{2}+m^{2}}$. A very deep antiproton-nucleus potential is obtained with the G-parity approach being $V_{opt}(\textbf{p}=0,\rho=\rho_{0}) = -655 $ MeV. From fitting the antiproton-nucleus scattering \cite{La09} and the real part of phenomenological antinucleon-nucleon optical potential \cite{Co82}, a factor $\xi$ is introduced in order to moderately evaluate the optical potential as $\Sigma_{S}^{\overline{N}}=\xi\Sigma_{S}^{N}$ and $\Sigma_{V}^{\overline{N}}=-\xi\Sigma_{V}^{N}$ with $\xi$=0.25, which leads to the strength of $V_{\overline{N}}=-164$ MeV at the normal nuclear density $\rho_{0}$=0.16 fm$^{-3}$.

Besides the reaction channels associated with resonances, hyperons and mesons in the model, the annihilation channels, charge-exchange reaction (CEX), elastic (EL) and inelastic scattering with antibaryons are included as follows \cite{Fe14}:
\begin{eqnarray}
&& \overline{B}B \rightarrow \texttt{annihilation}(\pi,\eta,\rho,\omega,K,\overline{K},\eta\prime,K^{\ast},\overline{K}^{\ast},\phi),
\nonumber  \\
&&  \overline{B}B \rightarrow \overline{B}B (\texttt{CEX, EL}),   \overline{N}N \leftrightarrow \overline{N}\Delta(\overline{\Delta}N), \overline{B}B \rightarrow \overline{Y}Y.
\nonumber  \\
\end{eqnarray}
Here the B strands for nucleon and $\Delta$(1232), Y($\Lambda$, $\Sigma$, $\Xi$), K(K$^{0}$, K$^{+}$) and $\overline{K}$($\overline{K^{0}}$, K$^{-}$). The overline of B (Y) means its antiparticle. The cross sections of these channels are based on the parametrization or extrapolation from available experimental data \cite{Bu12}. The annihilation dynamics in antibaryon-baryon collisions is described by a statistical model with SU(3) symmetry inclusion of all pseudoscalar and vector mesons \cite{Go92}, which considers various combinations of possible mesons with the final state from two to six particles \cite{La12}. Pions as the dominant products in the annihilation of antiproton in the nucleus contribute the energy deposition.

\section{Results and discussion}

The particle production in antiproton induced reactions is a significant probe in understanding the in-medium properties of particles, annihilation mechanism, strangeness exchange reactions etc. The localized energy is deposited in a target nucleus via the collisions of particles and nucleons, which makes the formation of highly excited nucleus. The particle emission is strongly influenced by surrounding nucleons. Shown in Fig. 1 is the mass dependence of pseudoscalar in antiproton induced reactions at an incident momentum of 200 MeV/c (kinetic energy of 21 MeV). The targets of $^{12}$C, $^{20}$Ne, $^{40}$Ca, $^{63}$Cu, $^{95}$Mo, $^{124}$Sn, $^{138}$Ba, $^{165}$Ho, $^{181}$Ta, $^{197}$Au and $^{238}$U are used for the bombardment with antiprotons. It is interest that the $\pi^{-}$ production increases with the mass of target nucleus, which is caused from the annihilation of antiprotons on neutrons. The secondary collisions of anti-kaons on nucleons reduces the anti-kaon yields, i.e., $\overline{K}N\rightarrow \pi Y$, which contribute the production of hyperons as shown in Fig. 2. The $\eta$ and kaons are insensitive to the target mass because of weakly interacting with nucleons. At the considered momentum below its threshold energy, e.g., the reaction $\overline{N}N\rightarrow \overline{\Lambda}\Lambda$ (p$_{threshold}$=1.439 GeV/c), hyperons are mainly contributed from the secondary collisions and strangeness exchange reactions after annihilations, i.e., $\pi N\rightarrow KY$ and $\overline{K}N\rightarrow \pi Y$.

\begin{figure*}
\includegraphics[width=16 cm]{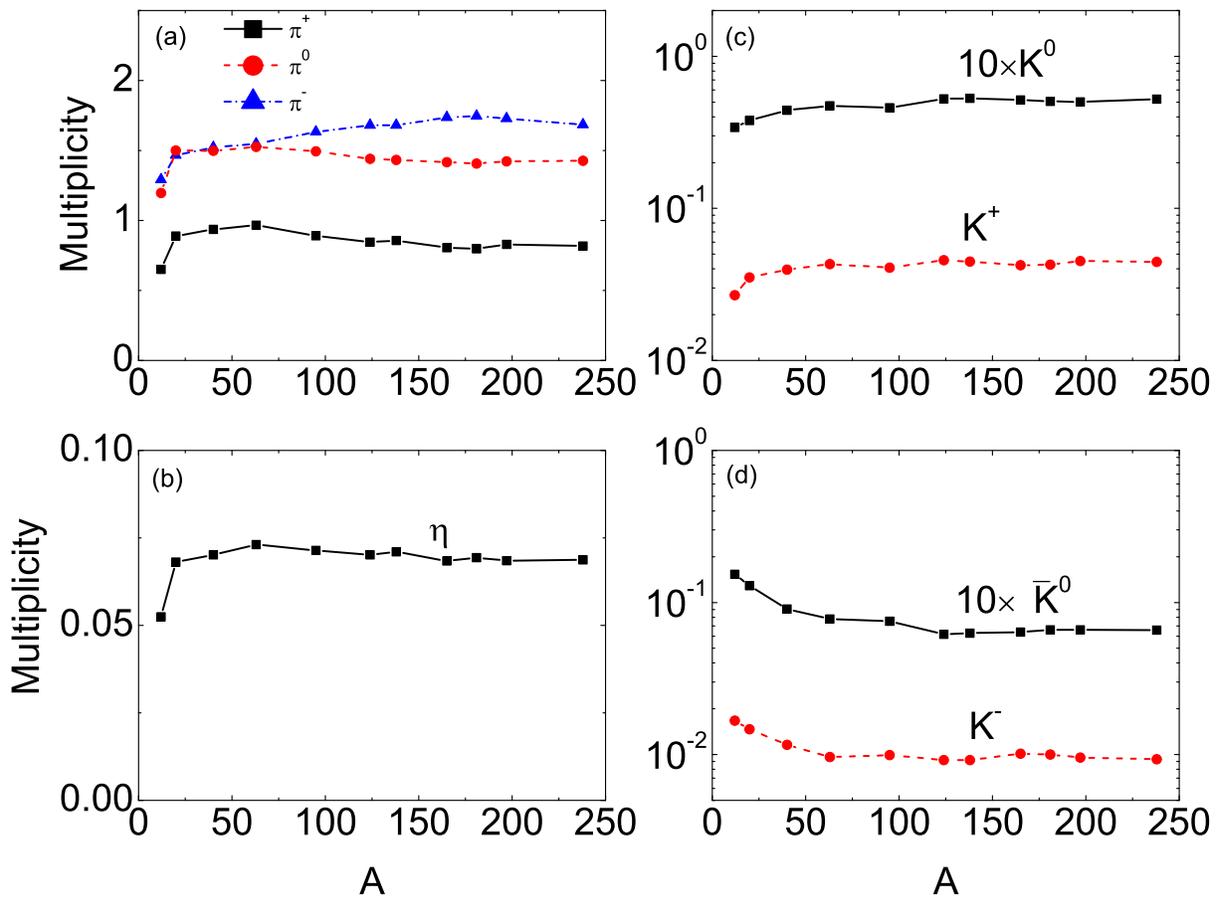}
\caption{\label{fig:wide} (Color online) Total multiplicities of pseudoscalar mesons (a) pions, (b) etas, (c) kaons and (d) antikaons as a function of mass number of target nuclei in antiproton induced reactions.}
\end{figure*}

\begin{figure*}
\includegraphics[width=16 cm]{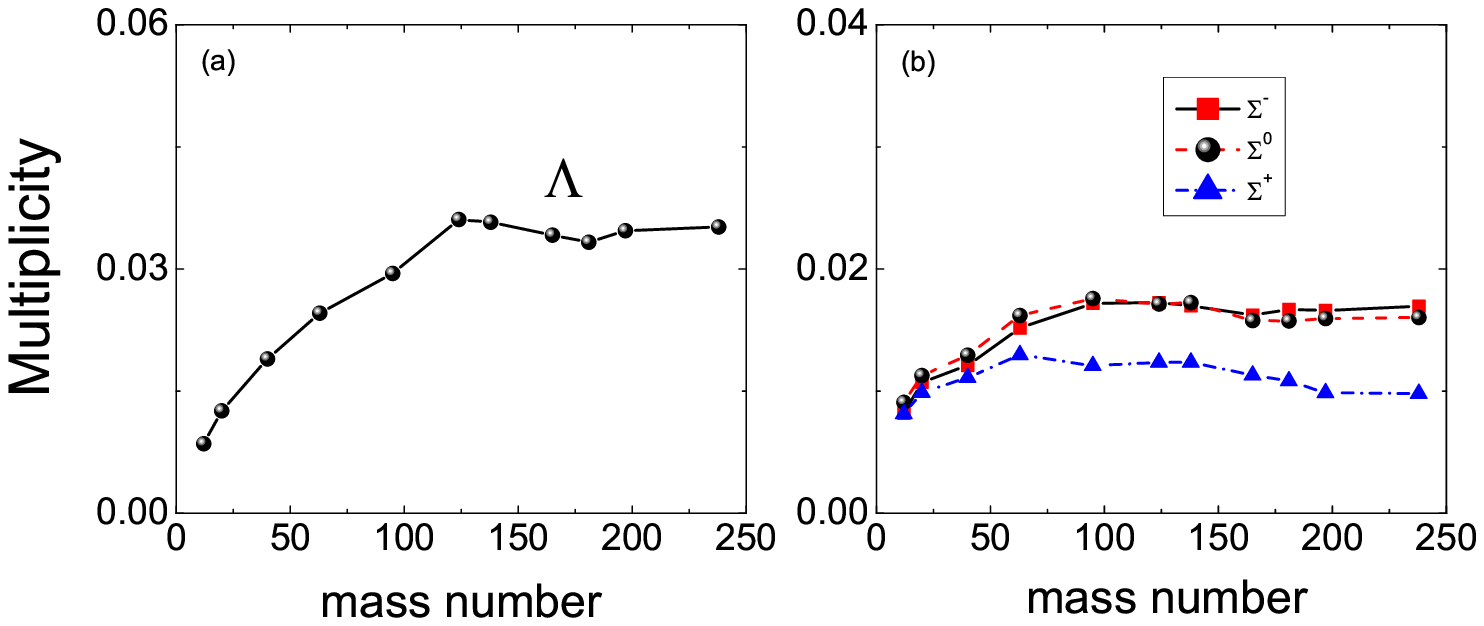}
\caption{\label{fig:wide} (Color online) The same as in Fig. 1, but for the (a) $\Lambda$ and (b) $\Sigma$ production.}
\end{figure*}

Besides the emission of mesons and hyperons after the annihilation of antiprotons in nuclei, target nuclei are excited via the collisions of particles and nucleons, which leads to evaporate nucleons and clusters from the transient nuclei and even to fragmentation reactions. The mass and charge yield distributions could be used to estimate the energy released by antiprotons in nuclei from the fragmentation magnitude. The nuclear dynamics induced by antiprotons is described by the LQMD model. The primary fragments are constructed in phase space with a coalescence model, in which nucleons at the freeze-out stage (equilibrium state for particle production) are considered to belong to one cluster with the relative momentum smaller than $P_{0}$ and with the relative distance smaller than $R_{0}$ (here $P_{0}$ = 200 MeV/c and $R_{0}$ = 3 fm). The primary fragments are highly excited. The de-excitation of the fragments is assumed to be isolated without rotation (zero angular momentum) and evaluated with the statistical code GEMINI \cite{Ch88}. The phase-space distributions of fragments manifest the excitation magnitude of target nucleus. The rapidity and kinetic energy spectra of light clusters ($Z \leq$2) are presented as shown in Fig. 3 and in Fig. 4, respectively. The shapes of different targets are very similar that mean the nuclear structure effects (separation energy, shell effect, pairing correlation etc) are negligible in the antiproton induced reactions.

\begin{figure*}
\includegraphics[width=16 cm]{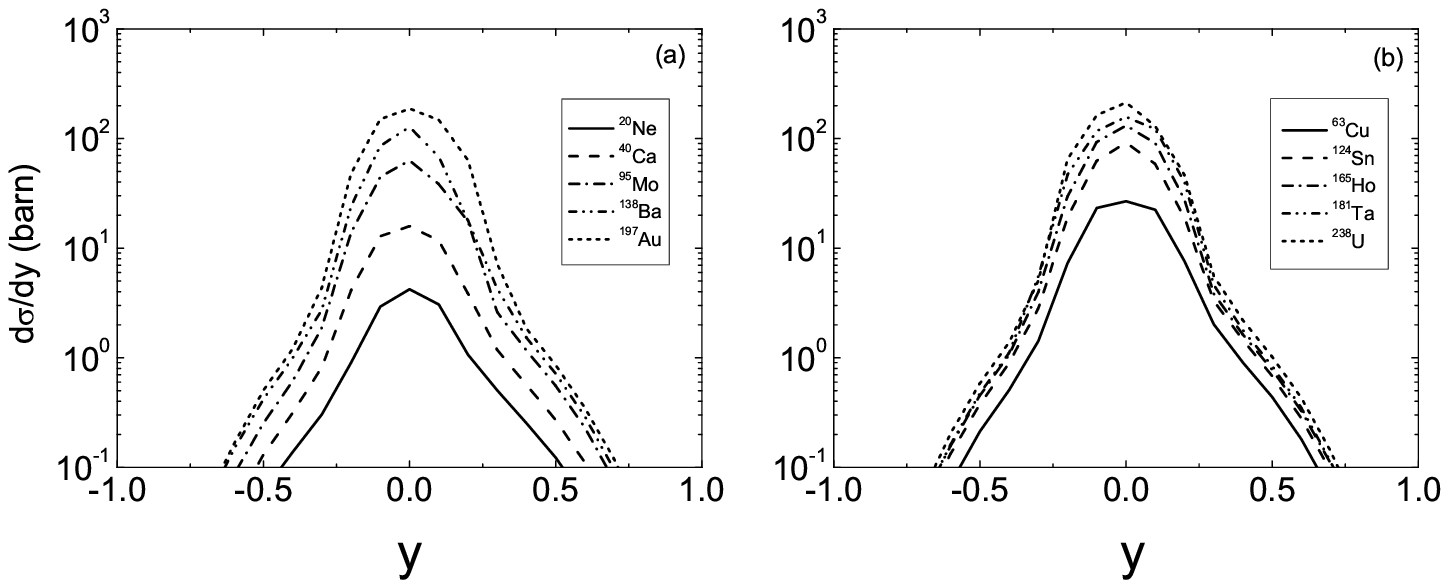}
\caption{\label{fig:wide} Rapidity distributions of light complex particles (nucleons, hydrogen and helium isotopes) in antiproton induced reactions on the nuclei of (a) $^{20}$Ne, $^{40}$Ca, $^{95}$Mo, $^{138}$Ba, $^{197}$Au and (b) $^{63}$Cu, $^{124}$Sn, $^{165}$Ho, $^{181}$Ta, $^{238}$U at the momentum of 200 MeV/c.}
\end{figure*}

\begin{figure*}
\includegraphics[width=16 cm]{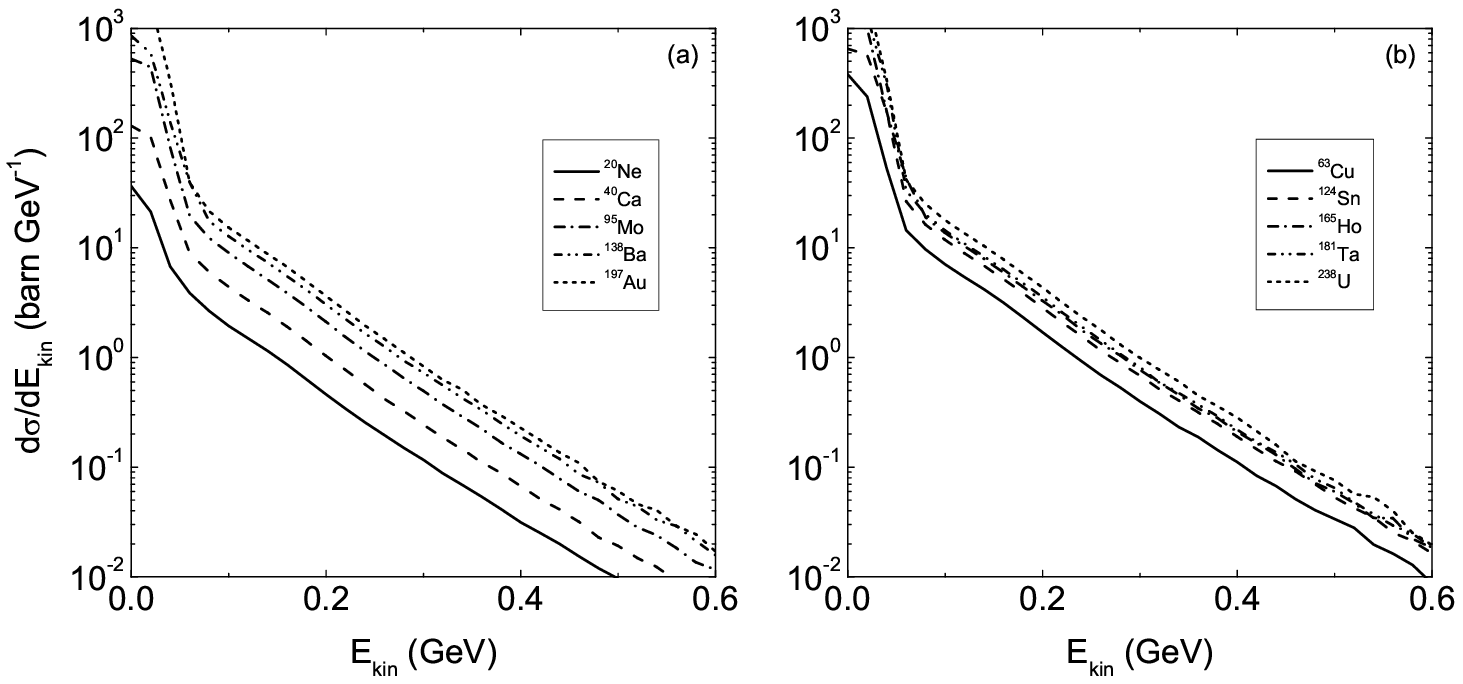}
\caption{\label{fig:wide} The same as in Fig. 3, but for the kinetic energy spectra.}
\end{figure*}

The fragmentation reactions induced by antiprotons were investigated in experiments at the LEAR at CERN. The energy deposition mechanism and properties of highly excited nuclei are concentrated on. The combined approach is tested in the fragmentation reactions of antiprotons on $^{197}$Au as shown in Fig. 5. The available data from the LEAR facility \cite{Lu02} could be nicely reproduced within the LQMD model combined with the GEMINI code. It is obvious that the primary fragments in the mass region of $100<A<160$ are highly excited and the yields are overestimated. The secondary decay leads to the appearance of bump structure around $A\sim 90$, which comes from the fission of heavy fragments. The minimum position of the yields appears at $A\sim 120$ and $Z\sim 50$ and the yields increase drastically with the fragments becoming heavier. Overall, the available data could be reproduced within the LQMD transport model combined with the GEMINI code. Moreover, based on the combined approach, we analyzed the fragmentation reactions with antiprotons at the incident momentum of 200 MeV/c on the light nuclei of $^{12}$C, $^{20}$Ne, $^{40}$Ca and $^{63}$Cu as shown in Fig. 6 and Fig. 7, respectively. A broad region of target-like nuclei is formed and the structure is similar, where the fragments are produced via the preequilibrium particle emission and evaporation from the hot nuclei. The fragmentation process can be understood as three stages, namely, antiproton-nucleon annihilation (energy released), meson-nucleon collisions (energy deposited) and fragmentation of highly excited nucleus (nuclear explosion).

\begin{figure*}
\includegraphics[width=16 cm]{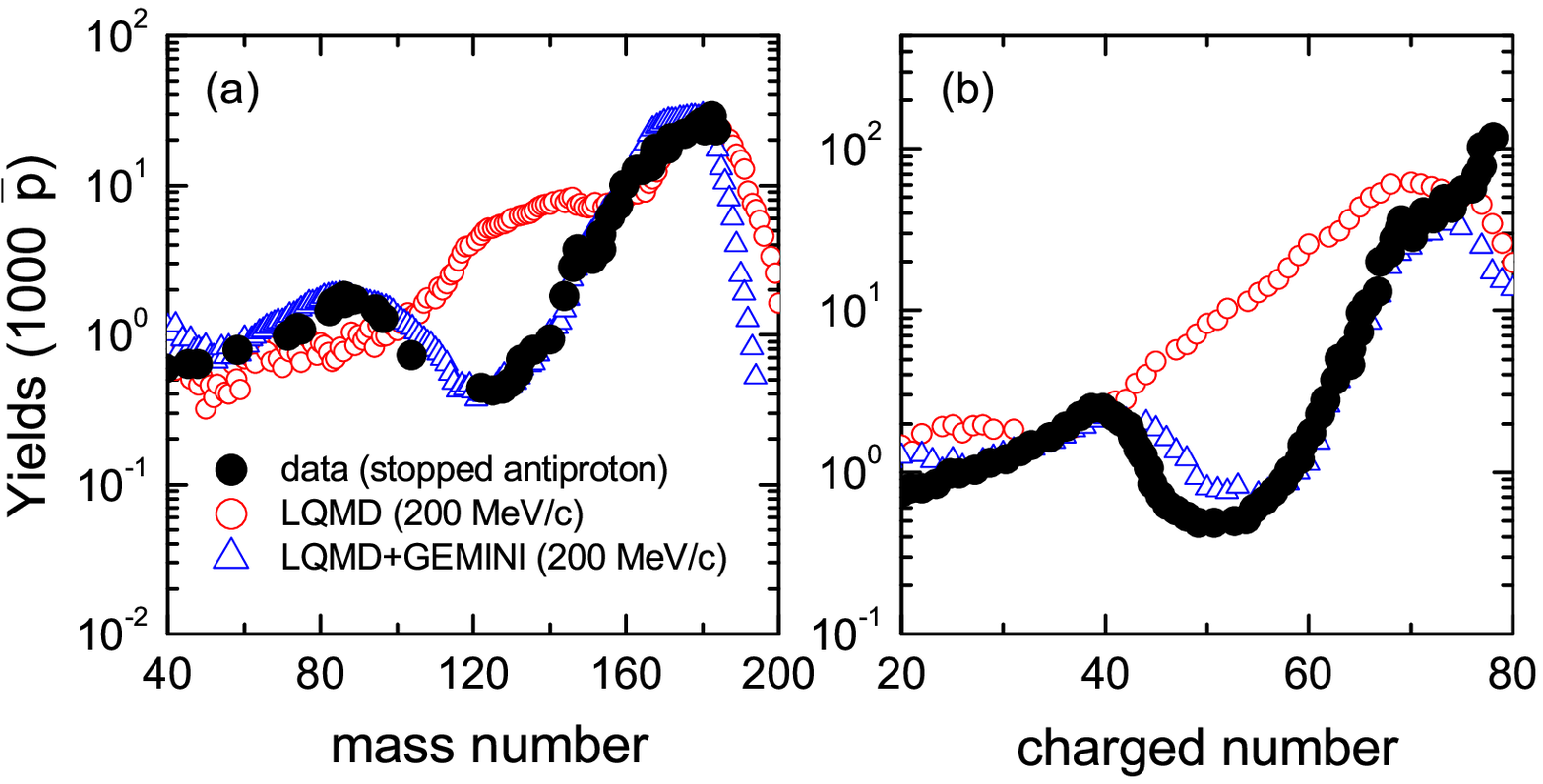}
\caption{\label{fig:wide} (a) Mass spectra and (b) charge distributions of fragments produced in the $\overline{p}$ + $^{197}$Au reaction at an incident momentum of 200 MeV/c combined with the statistical decay code GEMINI. The mass yields from LEAR facility at CERN \cite{Lu02} are shown for comparison.}
\end{figure*}

\begin{figure*}
\includegraphics[width=16 cm]{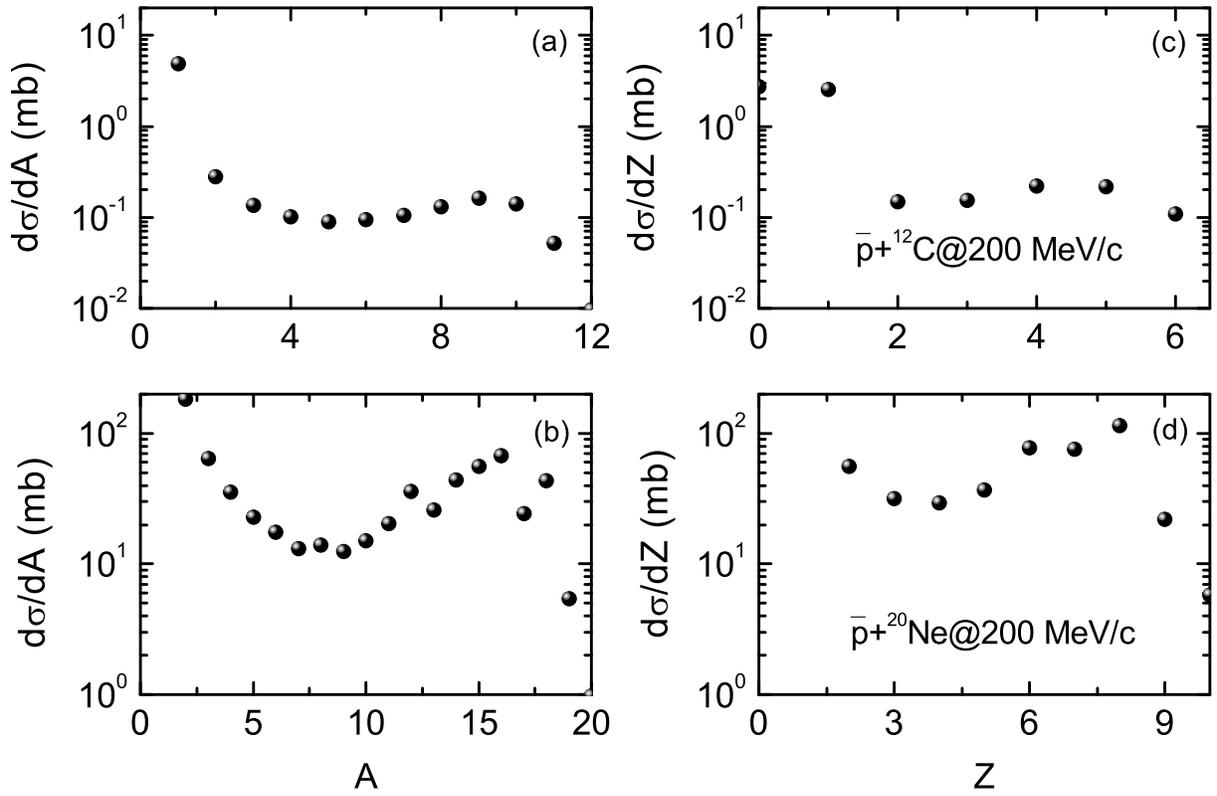}
\caption{\label{fig:wide} Fragment distributions in antiproton induced reactions at the incident momentum of 200 MeV/c as functions of (a) mass number on $^{12}$C, (b) mass number on $^{20}$Ne, (c) charged number on $^{12}$C, and (d) charged number on $^{20}$Ne, respectively.}
\end{figure*}

\begin{figure*}
\includegraphics[width=16 cm]{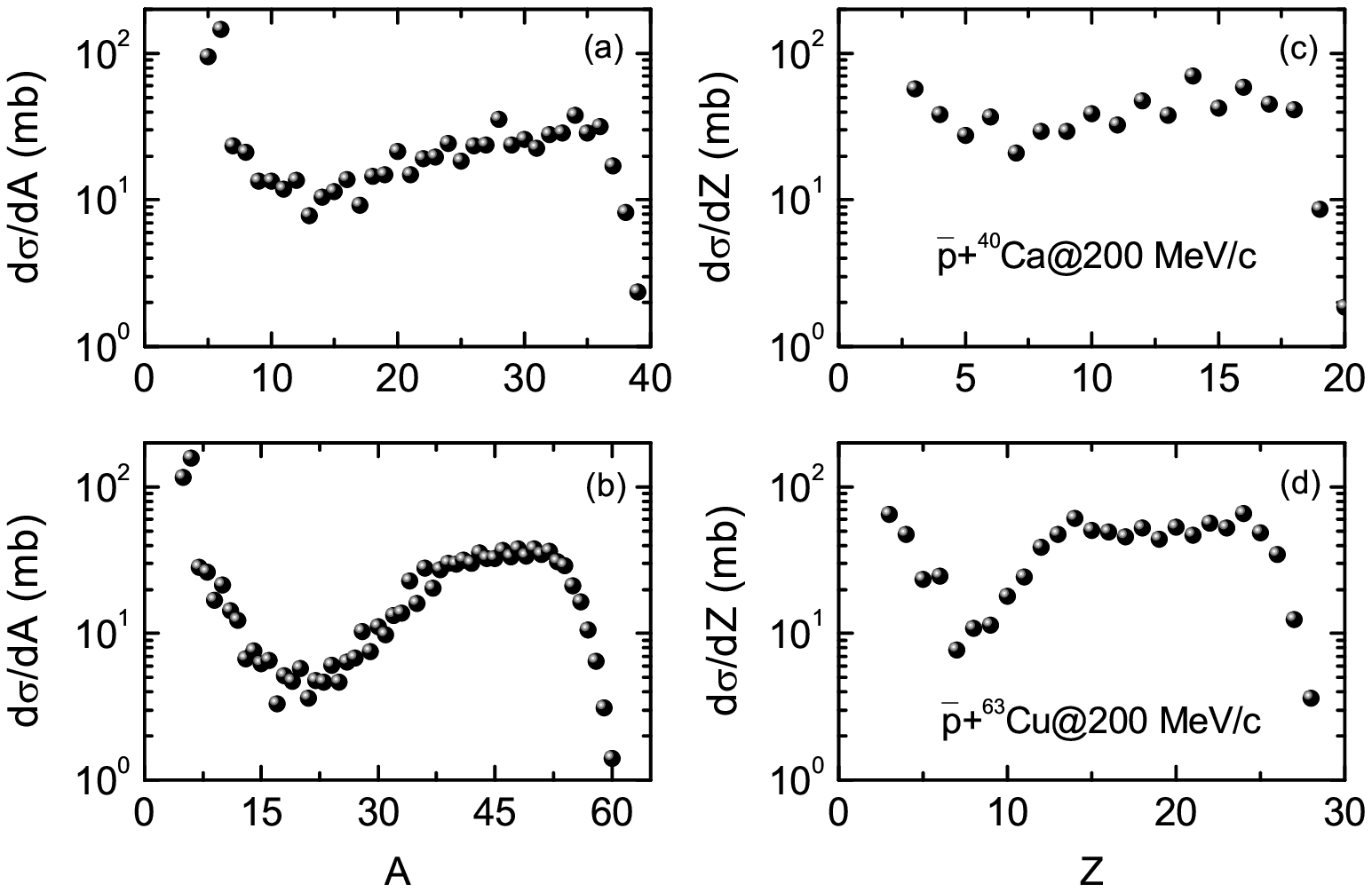}
\caption{\label{fig:wide} The same as in Fig. 6, but for (a) mass number on $^{40}$Ca, (b) mass number on $^{63}$Cu, (c) charged number on $^{40}$Ca, and (d) charged number on $^{63}$Cu, respectively.}
\end{figure*}

The fragmentation process of target nucleus induced by antiproton undergoes the explosive process (fast stage) in which the preequilibrium nucleons (light clusters) are emitted or the multifragments are produced after collisions between baryons and nucleons, and the decay process (slow stage) of the highly excited nucleus after the relative motion energy is deposited via the meson-nucleon collisions. The decay mechanism is determined by the excitation energy, i.e., the particle evaporation or fission dominating at low excitation energies (1-2 MeV/nucleon). The system is broken via multifragmentation when the local energy is close to the binding energy. More sophisticated investigations on the fragmentation reactions are performed. Shown in Fig. 8 is the mass (left window) and charge (right window) distributions of fragments and compared with the LEAR data with stopped antiprotons \cite{Mo89}. The production yields per antiproton are consistent with the available data. The average excitation energies are 157 MeV and 415 MeV for the target nuclei $^{95}$Mo and $^{165}$Ho, respectively, which are evaluated from the formula of $<E^{\ast}>=\int E^{\ast}\sigma(E^{\ast})dE^{\ast}/\int \sigma(E^{\ast})dE^{\ast}$, and $\sigma(E^{\ast})$ being the sum of cross sections of fragments at the excitation energy of $E^{\ast}$. The multifragmentation into intermediate mass fragments ($3<Z<30$) is negligible because of the limitation of excitation energy. Similar structures on the targets of $^{124}$Sn and $^{138}$Ba are found in Fig. 9 with the excitation energies of 238 MeV and 288 MeV, respectively. Roughly, the deposited energy increases with the mass of target nucleus, but weakly depends on the incident energy. Contribution of the fission fragments on the spectra is pronounced with heavy target as shown in Fig. 10, in particular for $^{238}$U.

\begin{figure*}
\includegraphics[width=16 cm]{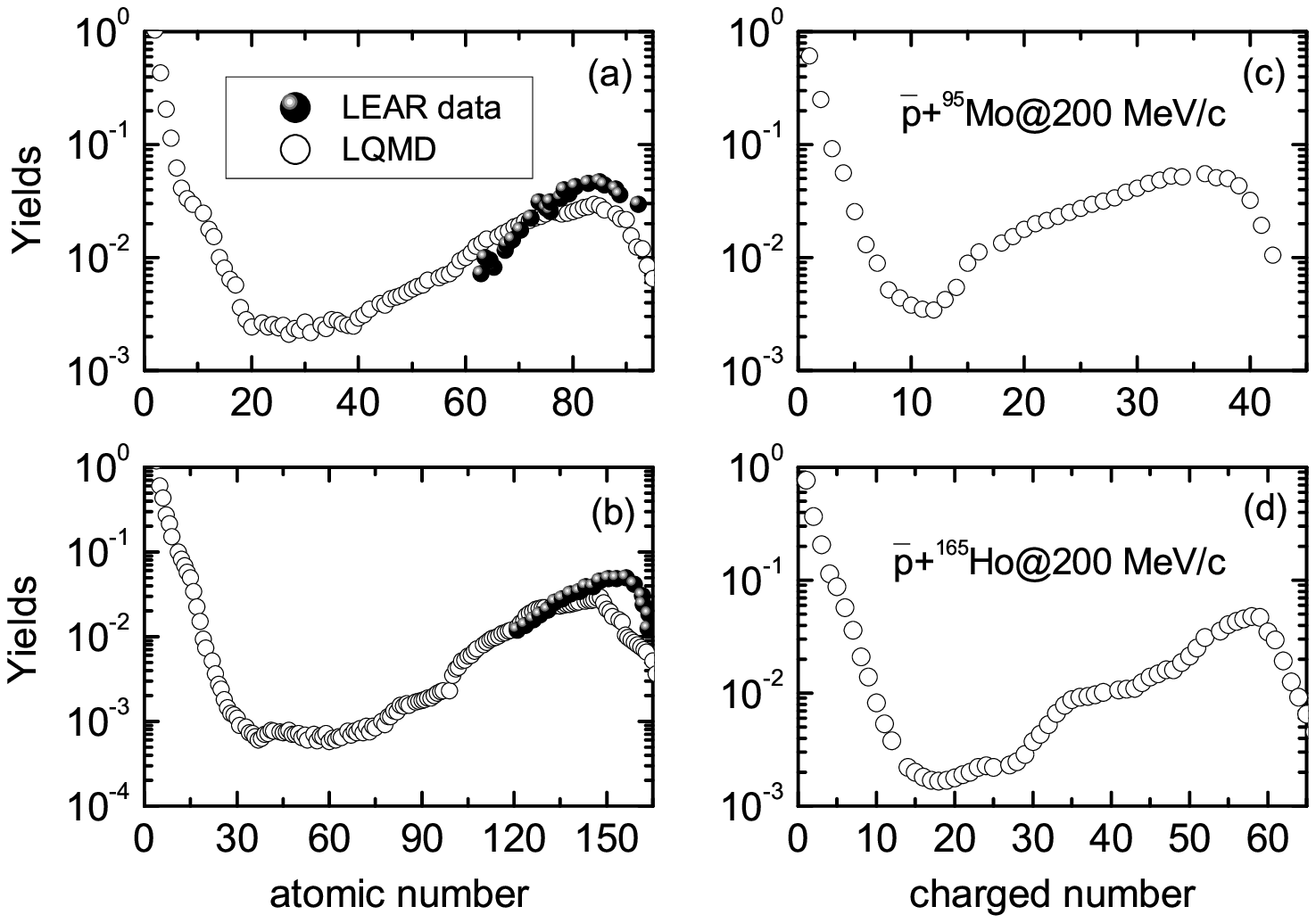}
\caption{\label{fig:wide} Fragmentation reactions induced by antiprotons and compared with the LEAR data \cite{Mo89} for the fragment (a) mass on $^{95}$Mo, (b) mass on $^{165}$Ho, (c) charged number $^{95}$Mo, and (d) charged number distributions on $^{165}$Ho, respectively.}
\end{figure*}

\begin{figure*}
\includegraphics[width=16 cm]{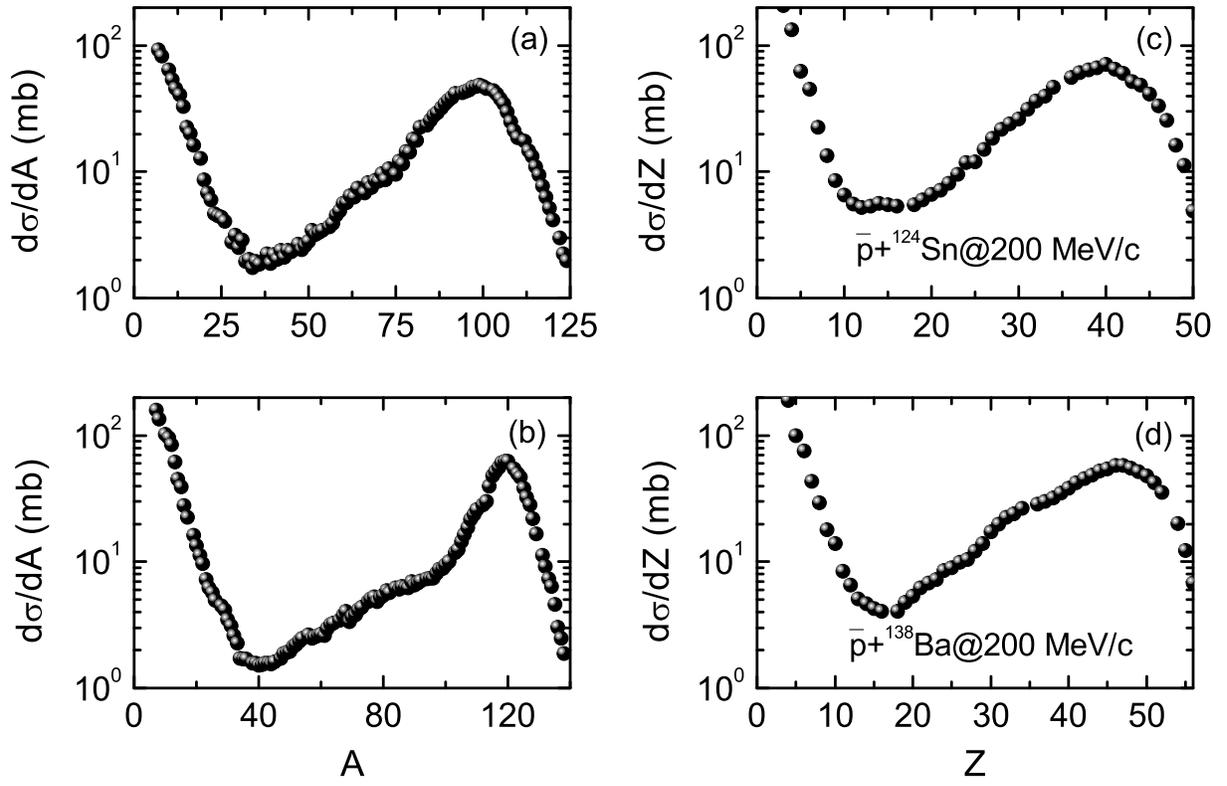}
\caption{\label{fig:wide} Fragment distributions in the antiprotons induced reactions for the spectra of mass (a) $^{124}$Sn, (b) $^{138}$Ba, charge number (c) $^{124}$Sn and (d) $^{138}$Ba, respectively.}
\end{figure*}

\begin{figure*}
\includegraphics[width=16 cm]{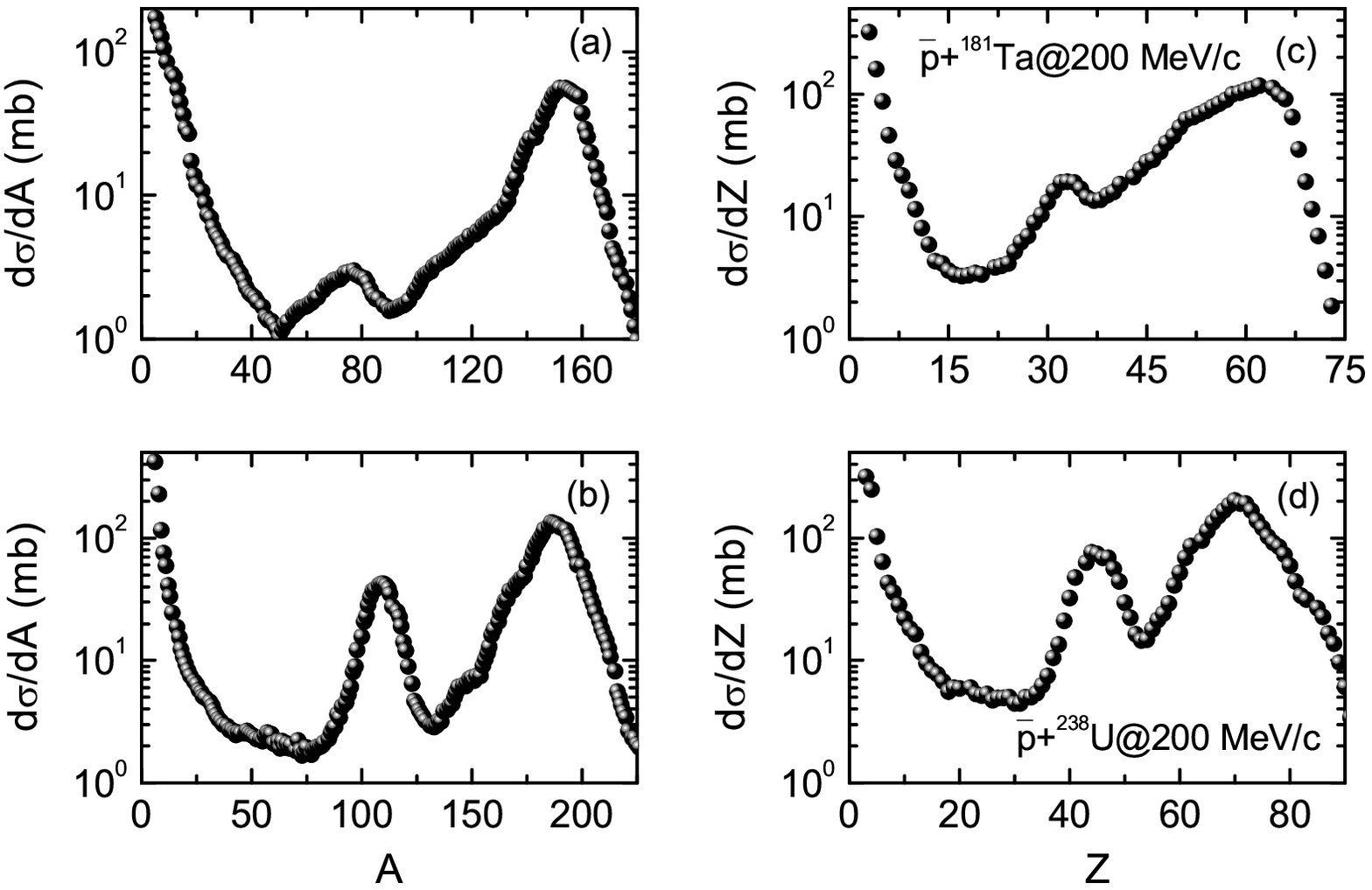}
\caption{\label{fig:wide} The same as in Fig. 9, but for the mass distributions on (a) $^{181}$Ta and (b) $^{238}$U as well as charge spectra on (c) $^{181}$Ta and (d) $^{238}$U, respectively.}
\end{figure*}

The mean nucleon removal in antiproton induced reactions manifests the meson-nucleon collision probability and is also related to the deposited energy in the nucleus. The average mass removals are evaluated from $\triangle A=A_{T}-1-\int_{A_{min}}^{A_{T}-2}\sigma(A)AdA/\int_{A_{min}}^{A_{T}-2}\sigma(A)dA$. Here, the $A_{T}$ and $A_{min}$ are the mass number of target nucleus and the integration limit being the first minimum position from the target-like fragments. Shown in Fig. 11 is a comparison of calculated removal nucleon number and the available data from the LEAR facility \cite{Lu02}. Reactions on the targets of $^{12}$C, $^{20}$Ne, $^{40}$Ca, $^{63}$Cu, $^{95}$Mo, $^{124}$Sn, $^{138}$Ba, $^{165}$Ho, $^{181}$Ta, $^{197}$Au and $^{238}$U with antiprotons at the incident momentum of 200 MeV/c are performed. The removal nucleons increase with the target nuclei because of the larger meson-nucleon collision probabilities for heavier nuclei. The averaged particle emission is related to the excitation energy. For example, on average 17.8 nucleons are emitted from the target $^{95}$Mo, and including 3.1 nucleons from the preequilibrium stage. Therefore, 14.7 nucleons evaporated from the thermal system and assuming the 8 MeV separation energy and 3 MeV kinetic energy per nucleon \cite{Po95} leads to the excitation energy of 161.7 MeV. The value is close to the evaluation from the energy spectra (157 MeV).

\begin{figure}
\includegraphics[width=8 cm]{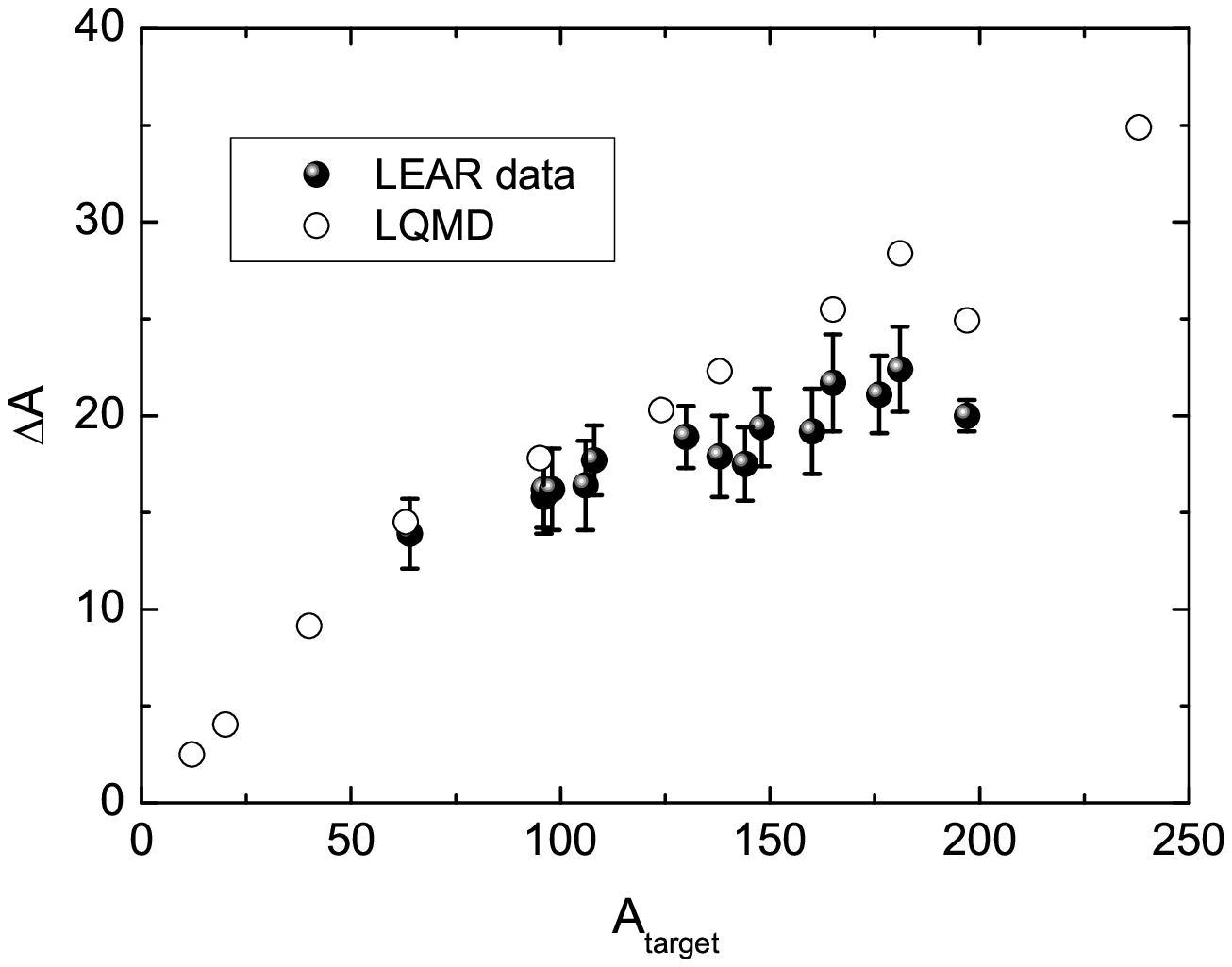}
\caption{\label{fig:epsart} Averaged mass removal as a function of mass number of target nuclei \cite{Lu02}.}
\end{figure}

The intermediate mass fragments (IMFs) produced in Fermi-energy heavy-ion collisions has been investigated for extracting the nuclear equation of state, liquid-gas phase transition and density dependence of symmetry energy \cite{Li04,Fe16}, where the composite system is formed at excitation energies of 10-20 MeV per nucleon. The fluctuation and explosive decay of the excited system contribute the IMF production. Shown in Fig. 12 is the correlation of the IMF multiplicity and charged particles on the targets of $^{95}$Mo, $^{124}$Sn, $^{165}$Ho, $^{197}$Au and $^{238}$U with antiprotons at the incident momentum of 200 MeV/c. A broad distribution is found for the heavier target. The IMF production in antiproton induced reactions is strongly compressed in comparison to heavy-ion collisions owing to the less fluctuation and lower excitation energy.

\begin{figure}
\includegraphics[width=8 cm]{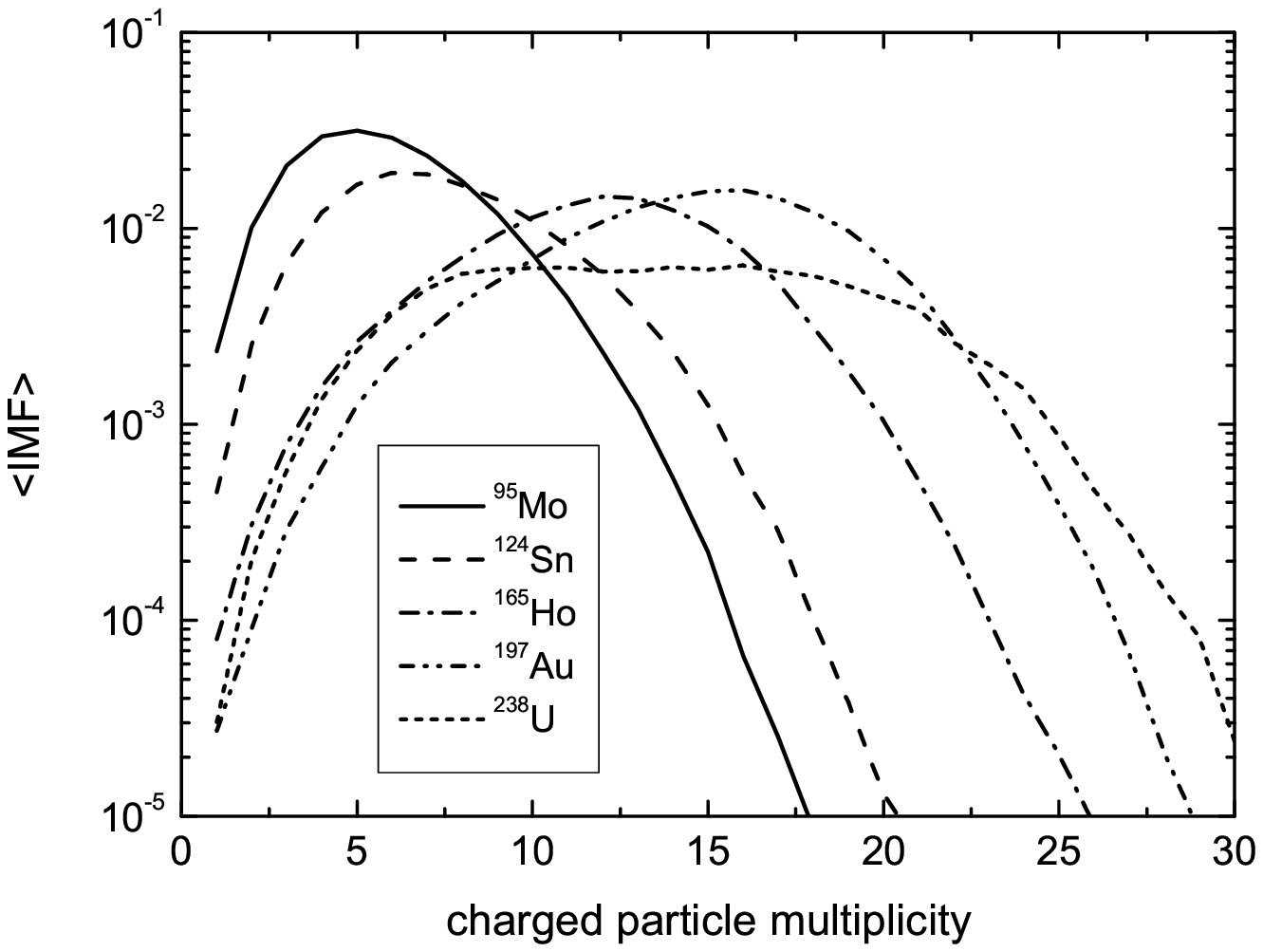}
\caption{\label{fig:epsart} Correlation of the intermediate mass fragments (IMFs) and charged particle multiplicity produced in $\overline{p}$ induced reactions on different targets.}
\end{figure}

\section{Conclusions}

The fragmentation reactions induced by low-energy antiprotons has been investigated within the LQMD transport model. The de-excitation of the primary fragments is described with the help of the statistical code GEMINI. The available data of the fragment production from the LEAR at CERN can be nicely reproduced with the combined approach. The energy released in the antiproton-nucleon annihilation is mainly deposited in the target nucleus via the pion-nucleon collisions. The averaged nucleons removed from the target nucleus increases with the mass number. The bump structure contributed from the fission fragments of heavy nuclei is observed. Hyperons are mainly produced via strangeness exchange reactions in collisions of antikaons and nucleons, which have smaller relative momentum and could be easily captured by the residue nuclei to form hypernuclei. The approach will provide a cornerstone for the antiproton physics at PANDA (Antiproton Annihilation at Darmstadt, Germany) in the near future, i.e., hypernuclei, in-medium properties of hadrons etc.

\bigskip
\textbf{Acknowledgements}

This work was supported by the Major State Basic Research Development Program in China (No. 2015CB856903), and the National Natural Science Foundation of China (No. 11675226 and No. 11175218).

\end{document}